\documentstyle[sprocl,epsfig,psfig,aps]{revtex}

\input{psfig}

\def\ra{\rightarrow}

\def\be{\begin{equation}}
\def\ee{\end{equation}}
\def\bea{\begin{eqnarray}}
\def\eea{\end{eqnarray}}

\newcommand{\newc}{\newcommand} 
\newc{\lra}{\leftrightarrow} 
\newc{\beq}{\begin{equation}} 
\newc{\eeq}{\end{equation}} 
\newc{\barr}{\begin{eqnarray}} 
\newc{\earr}{\end{eqnarray}} 

\begin{document} 
\title{Theoretical Rates for Direct Detection of SUSY Dark Matter}

\author{J. D. VERGADOS} 

\address{Theoretical Physics Division, University of Ioannina, GR-45110, 
Greece\\E-mail:Vergados@cc.uoi.gr} 

\maketitle\abstracts{
Exotic dark matter together with the vacuum energy (associated with the 
cosmological constant) seem to dominate in the Universe.
Thus its direct detection 
is central to particle physics and cosmology. Supersymmetry
provides a natural dark matter candidate, the lightest
supersymmetric particle (LSP). Furthermore from the knowledge of the density
 and velocity distribution of the LSP, 
the quark substructure of the nucleon and the nuclear structure (form factor 
and/or spin response function), one is able to evaluate
the event rate for LSP-nucleus elastic scattering. The thus obtained event rates
are, however, very low.
 So it is imperative to exploit the modulation effect, i.e. the dependence of
the event rate on  the Earth's motion and the directional signature, i.e.
the dependence of the rate on the direction of the recoiling
nucleus. In this paper we study such experimental signatures
 employing a supersymmetric model with universal boundary conditions
at large $tan\beta$.
}

\date{\today}

\maketitle

\section{Introduction}
In recent years the consideration of exotic dark matter has become necessary
in order to close the Universe \cite{Jungm}.
 The COBE data ~\cite{COBE} suggest that CDM (Cold Dark Matter)
is at least $60\%$ ~\cite {GAW}. On the other hand 
evidence from two different teams,
the High-z Supernova Search Team \cite {HSST} and the
Supernova Cosmology Project  ~\cite {SPF} $^,$~\cite {SCP} 
  suggests that the Universe may be dominated by 
the  cosmological constant $\Lambda$.
Thus the situation can be adequately
described by  a baryonic component $\Omega_B=0.1$ along with the exotic 
components $\Omega _{CDM}= 0.3$ and $\Omega _{\Lambda}= 0.6$
(see next section for the definitions).
In another analysis Turner \cite {Turner} gives 
$\Omega_{m}= \Omega _{CDM}+ \Omega _B=0.4$.
Since the non exotic component cannot exceed $40\%$ of the CDM 
~\cite{Jungm},~\cite {Benne}, there is room for the exotic WIMP's 
(Weakly  Interacting Massive Particles).
  In fact the DAMA experiment ~\cite {BERNA2} 
has claimed the observation of one signal in direct detection of a WIMP, which
with better statistics has subsequently been interpreted as a modulation signal
~\cite{BERNA1}.

 In the most favored scenario of supersymmetry the
LSP can be simply described as a Majorana fermion, a linear 
combination of the neutral components of the gauginos and Higgsinos
\cite{Jungm,ref1,Gomez,ref2}.

\section{An Overview of Direct Detection - The Allowed SUSY Parameter Space.}

 Since this particle is expected to be very massive, $m_{\chi} \geq 30 GeV$, and
extremely non relativistic with average kinetic energy $T \leq 100 KeV$,
it can be directly detected ~\cite{JDV96,KVprd} mainly via the recoiling
of a nucleus (A,Z) in elastic scattering.
In order to compute the event rate one needs
the following ingredients:

1) An effective Lagrangian at the elementary particle 
(quark) level obtained in the framework of supersymmetry as described 
, e.g., in Refs.~\cite{Jungm,ref2}.

2) A procedure in going from the quark to the nucleon level, i.e. a quark 
model for the nucleon. The results depend crucially on the content of the
nucleon in quarks other than u and d. This is particularly true for the scalar
couplings as well as the isoscalar axial coupling ~\cite{Dree}$^-$\cite{Chen}.

3) Compute the relevant nuclear matrix elements \cite{KVdubna,DIVA00}
using as reliable as possible many body nuclear wave functions.
The situation is a bit simpler in the case of the scalar coupling, in which
case one only needs the nuclear form factor.

Since the obtained rates are very low, one would like to be able to exploit the
modulation of the event rates due to the earth's
revolution around the sun \cite{Verg98,Verg99}$^-$\cite{Verg01}. To this end one
adopts a folding procedure
assuming some distribution~\cite{Jungm,Verg99,Verg01} of velocities for the LSP.
One also would like to know the directional rates, by observing the 
nucleus in a certain direction, which correlate with the motion of the
sun around the center of the galaxy and the motion of the Earth 
\cite {ref1,UKDMC}.

 The calculation of this cross section  has become pretty standard.
 One starts with   
representative input in the restricted SUSY parameter space as described in
the literature~\cite{Gomez,ref2}. 
We will adopt a phenomelogical procedure taking  universal soft 
SUSY breaking terms at $M_{GUT}$, i.e., a 
common mass for all scalar fields $m_0$, a common gaugino mass 
$M_{1/2}$ and a common trilinear scalar coupling $A_0$, which 
we put equal to zero (we will discuss later the influence of 
non-zero $A_0$'s). Our effective theory below $M_{GUT}$ then 
depends on the parameters \cite{Gomez}:
\[
m_0,\ M_{1/2},\ \mu_0,\ \alpha_G,\ M_{GUT},\ h_{t},\ ,\ h_{b},\ ,\ h_{\tau},\ 
\tan\beta~,  
\]
where $\alpha_G=g_G^2/4\pi$ ($g_G$ being the GUT gauge coupling 
constant) and $h_t, h_b, h_\tau $ are respectively the top, bottom and 
tau Yukawa coupling constants at $M_{GUT}$. The values of $\alpha_G$ and 
$M_{GUT}$ are obtained as described in Ref.\cite{Gomez}.
For a specified value of $\tan\beta$ at $M_S$, we determine $h_{t}$ at 
$M_{GUT}$ by fixing the top quark mass at the center of its 
experimental range, $m_t(m_t)= 166 \rm{GeV}$. The value
of  $h_{\tau}$ at $M_{GUT}$ is  fixed by using the running tau lepton
mass at $m_Z$, $m_\tau(m_Z)= 1.746 \rm{GeV}$. 
The value of $h_{b}$ at $M_{GUT}$ used is such that:
\[
m_b(m_Z)_{SM}^{\overline{DR}}=2.90\pm 0.14~{\rm GeV}.
\]
after including the SUSY threshold correction. 
The SUSY parameter space is subject to the
 following constraints:\\
1.) The LSP relic abundance will satisfy the cosmological constrain:
\begin{equation}
0.09 \le \Omega_{LSP} h^2 \le 0.22
\label{eq:in2}
\end{equation}
 2.) The Higgs bound obtained from recent CDF \cite{VALLS}
 and LEP2 \cite{DORMAN},
 i.e. $m_h~>~113~GeV$.\\
3.) We will limit ourselves to LSP-nucleon cross sections for the scalar
coupling, which gives detectable rates
\begin{equation}
4\times 10^{-7}~pb~ \le \sigma^{nucleon}_{scalar} 
\le 2 \times 10^{-5}~pb~
\label{eq:in3}
\end{equation}
 We should remember that the event rate does not depend only
 on the nucleon cross section, but on other parameters also, mainly
 on the LSP mass and the nucleus used in target. 
The condition on the nucleon cross section imposes severe constraints on the
acceptable parameter space. In particular in our model it restricts 
$tan \beta$ to values $tan \beta \simeq 50$. We will not elaborate further
on this point, since it has already appeared \cite{JDV02}. 
\bigskip
\section{Expressions for the Differential Cross Section .} 
\bigskip

 The effective Lagrangian describing the LSP-nucleus cross section can
be cast in the form \cite {JDV96}
 \beq
{\it L}_{eff} = - \frac {G_F}{\sqrt 2} \{({\bar \chi}_1 \gamma^{\lambda}
\gamma_5 \chi_1) J_{\lambda} + ({\bar \chi}_1 
 \chi_1) J\}
 \label{eq:eg 41}
\eeq
where
\beq
  J_{\lambda} =  {\bar N} \gamma_{\lambda} (f^0_V +f^1_V \tau_3
+ f^0_A\gamma_5 + f^1_A\gamma_5 \tau_3)N~~,~~
J = {\bar N} (f^0_s +f^1_s \tau_3) N
 \label{eq:eg.42}
\eeq

We have neglected the uninteresting pseudoscalar and tensor
currents. Note that, due to the Majorana nature of the LSP, 
${\bar \chi_1} \gamma^{\lambda} \chi_1 =0$ (identically).

 With the above ingredients the differential cross section can be cast in the 
form \cite{ref1,Verg98,Verg99}
\begin{equation}
d\sigma (u,\upsilon)= \frac{du}{2 (\mu _r b\upsilon )^2} [(\bar{\Sigma} _{S} 
                   +\bar{\Sigma} _{V}~ \frac{\upsilon^2}{c^2})~F^2(u)
                       +\bar{\Sigma} _{spin} F_{11}(u)]
\label{2.9}
\end{equation}
\begin{equation}
\bar{\Sigma} _{S} = \sigma_0 (\frac{\mu_r(A)}{\mu _r(N)})^2  \,
 \{ A^2 \, [ (f^0_S - f^1_S \frac{A-2 Z}{A})^2 \, ] \simeq \sigma^S_{p,\chi^0}
        A^2 (\frac{\mu_r(A)}{\mu _r(N)})^2 
\label{2.10}
\end{equation}
\begin{equation}
\bar{\Sigma} _{spin}  =  \sigma^{spin}_{p,\chi^0}~\zeta_{spin}~~,~~
\zeta_{spin}= \frac{(\mu_r(A)/\mu _r(N))^2}{3(1+\frac{f^0_A}{f^1_A})^2}S(u)
\label{2.10a}
\end{equation}
\begin{equation}
S(u)=[(\frac{f^0_A}{f^1_A} \Omega_0(0))^2 \frac{F_{00}(u)}{F_{11}(u)}
  +  2\frac{f^0_A}{ f^1_A} \Omega_0(0) \Omega_1(0)
\frac{F_{01}(u)}{F_{11}(u)}+  \Omega_1(0))^2  \, ] 
\label{2.10b2}
\end{equation}
\begin{equation}
\bar{\Sigma} _{V}  =  \sigma^V_{p,\chi^0}~\zeta_V 
\label{2.10c}
\end{equation}
\begin{equation}
\zeta_V =  \frac{(\mu_r(A)/\mu _r(N))^2}{(1+\frac{f^1_V}{f^0_V})^2} A^2 \, 
(1-\frac{f^1_V}{f^0_V}~\frac{A-2 Z}{A})^2 [ (\frac{\upsilon_0} {c})^2  
[ 1  -\frac{1}{(2 \mu _r b)^2} \frac{2\eta +1}{(1+\eta)^2} 
\frac{\langle~2u~ \rangle}{\langle~\upsilon ^2~\rangle}] 
\label{2.10d}
\end{equation}
\\
$\sigma^i_{p,\chi^0}=$ proton cross-section,$i=S,spin,V$ given by:\\
$\sigma^S_{p,\chi^0}= \sigma_0 ~(f^0_S)^2~(\frac{\mu _r(N)}{m_N})^2$ 
 (scalar) , 
(the isovector scalar is negligible, i.e. $\sigma_p^S=\sigma_n^S)$\\
$\sigma^{spin}_{p,\chi^0}=\sigma_0~~3~(f^0_A+f^1_A)^2~(\frac{\mu _r(N)}{m_N})^2$ 
  (spin) ,
$\sigma^{V}_{p,\chi^0}= \sigma_0~(f^0_V+f^1_V)^2~(\frac{\mu _r(N)}{m_N})^2$ 
(vector)   \\
where $m_N$ is the nucleon mass,
 $\eta = m_x/m_N A$, and
 $\mu_r(A)$ is the LSP-nucleus reduced mass,  
 $\mu_r(N)$ is the LSP-nucleon reduced mass and  
\begin{equation}
\sigma_0 = \frac{1}{2\pi} (G_F m_N)^2 \simeq 0.77 \times 10^{-38}cm^2 
\label{2.7} 
\end{equation}
\begin{equation}
Q=Q_{0}u~~, \qquad Q_{0} = \frac{1}{A m_{N} b^2}=4.1\times 10^4A^{-4/3}~KeV 
\label{2.15} 
\end{equation}
where
Q is the energy transfer to the nucleus and
$F(u)$ is the nuclear form factor. 

In the present paper we will concentrate on the coherent mode. For a discussion
of the spin contribution, expected to be important in the case of the light
nuclei, has been reviewed elsewhere \cite{JDV02}.

\section{Expressions for the Rates.} 
 The non-directional event rate is given by:
\begin{equation}
R=R_{non-dir} =\frac{dN}{dt} =\frac{\rho (0)}{m_{\chi}} \frac{m}{A m_N} 
\sigma (u,\upsilon) | {\boldmath \upsilon}|
\label{2.17} 
\end{equation}
 Where
 $\rho (0) = 0.3 GeV/cm^3$ is the LSP density in our vicinity and 
 m is the detector mass 
The differential non-directional  rate can be written as
\begin{equation}
dR=dR_{non-dir} = \frac{\rho (0)}{m_{\chi}} \frac{m}{A m_N} 
d\sigma (u,\upsilon) | {\boldmath \upsilon}|
\label{2.18}  
\end{equation}
where $d\sigma(u,\upsilon )$ was given above.

 The directional differential rate \cite{ref1},\cite{Verg01} in the
direction $\hat{e}$ is given by :
\begin{equation}
dR_{dir} = \frac{\rho (0)}{m_{\chi}} \frac{m}{A m_N} 
{\boldmath \upsilon}.\hat{e} H({\boldmath \upsilon}.\hat{e})
 ~\frac{1}{2 \pi}~  
d\sigma (u,\upsilon)
\label{2.20}  
\end{equation}
where H the Heaviside step function. The factor of $1/2 \pi$ is 
introduced, since  the differential cross section of the last equation
is the same with that entering the non-directional rate, i.e. after
an integration
over the azimuthal angle around the nuclear momentum has been performed.
In other words, crudely speaking, $1/(2 \pi)$ is the suppression factor we
 expect in the directional rate compared to the usual one. The precise 
suppression factor depends, of course, on the direction of observation.
The mean value of the non-directional event rate of Eq. (\ref {2.18}), 
is obtained by convoluting the above expressions with the LSP velocity
distribution $f({\bf \upsilon}, {\boldmath \upsilon}_E)$ 
with respect to the Earth, i.e. is given by:
\beq
\Big<\frac{dR}{du}\Big> =\frac{\rho (0)}{m_{\chi}} 
\frac{m}{A m_N}  
\int f({\bf \upsilon}, {\boldmath \upsilon}_E) 
          | {\boldmath \upsilon}|
                       \frac{d\sigma (u,\upsilon )}{du} d^3 {\boldmath \upsilon} 
\label{3.10} 
\eeq
 The above expression can be more conveniently written as
\beq
\Big<\frac{dR}{du}\Big> =\frac{\rho (0)}{m_{\chi}} \frac{m}{Am_N} \sqrt{\langle
\upsilon^2\rangle } {\langle \frac{d\Sigma}{du}\rangle }~,~ 
\langle \frac{d\Sigma}{du}\rangle =\int
           \frac{   |{\boldmath \upsilon}|}
{\sqrt{ \langle \upsilon^2 \rangle}} f({\boldmath \upsilon}, 
         {\boldmath \upsilon}_E)
                       \frac{d\sigma (u,\upsilon )}{du} d^3 {\boldmath \upsilon}
\label{3.11}  
\eeq

 After performing the needed integrations over the velocity distribution,
to first order in the Earth's velocity, and over the energy transfer u  the
 last expression takes the form
\beq
R =  \bar{R}~t~
          [1 + h(a,Q_{min})cos{\alpha})] 
\label{3.55a}  
\eeq
where $\alpha$ is the phase of the Earth ($\alpha=0$ around June 2nd)
and  $Q_{min}$ is the energy transfer cutoff imposed by the detector.
In the above expressions $\bar{R}$ is the rate obtained in the conventional 
approach \cite {JDV96} by neglecting the folding with the LSP velocity and the
momentum transfer dependence of the differential cross section, i.e. by
\beq
\bar{R} =\frac{\rho (0)}{m_{\chi}} \frac{m}{Am_N} \sqrt{\langle
v^2\rangle } [\bar{\Sigma}_{S}+ \bar{\Sigma} _{spin} + 
\frac{\langle \upsilon ^2 \rangle}{c^2} \bar{\Sigma} _{V}]
\label{3.39b}  
\eeq
where $\bar{\Sigma} _{i}, i=S,V,spin$ 
 contain all the parameters of the
SUSY models. 
 The modulation is described by the parameter $h$ .

The total  directional event rates  can be obtained in a similar fashion by
by integrating Eq. (\ref {2.20}) 
with respect to the velocity as well as the energy transfer u.  We find

\beq
R_{dir}  =   \bar{R} [(t_{dir}/2 \pi) \, 
            [1 + (h_1-h_2)cos{\alpha}) + h_3 sin{\alpha}]
\label{4.56}  
\eeq
where the quantity $t_{dir}$ provides the un modulated amplitude, while $h_1,h_2$
 and $h_3$ describe the modulation. They are functions of the angles 
$\Theta$ and $\Phi$, which specify the direction of observation $\hat{e}$,
  as well as the
parameters $a$ and $Q_{min}$.
 The effect of folding
with LSP velocity on the total rate is taken into account via the quantity
$t_{dir}$, which depends on the LSP mass. All other SUSY parameters have been
 absorbed in $\bar{R}$. 
 In the special case previously studied, i.e 
along the coordinate axes, we find that: a) in the direction of the sun's
 motion
 $h_2=h_3=0$, b)
along the radial direction (y axes) $h_3=0$ and c) in the vertical to the
 galaxy $h_2=0$.
 Instead of $t_{dir}$ itself it is more convenient
to present the reduction factor of the un modulated directional rate compared
to the usual non-directional one, i.e.
\beq
f_{red}=\frac{R_{dir}}{R}=t_{dir}/(2 \pi~t)= \kappa/(2 \pi) 
\label{kappa.eq}
\eeq
  It turns out that the parameter $\kappa$, being the 
ratio of two rates, is less dependent on these parameters.
Given the functions $h_l(a,Q_{min})$, $l=1,2,3$, one can plot the the
expression in
Eqs (\ref {3.55a}) and \ref{4.56} as a function of the phase of the earth
 $\alpha$. 
\section{The Scalar Contribution- The Role of the Heavy Quarks}
\bigskip

The coherent scattering can be mediated via the the neutral
intermediate Higgs particles (h and H), which survive as physical 
particles. It can also be mediated via s-quarks, via the mixing
of the isodoublet and isosinlet s-quarks of the same charge. In our
model we find that the Higgs contribution becomes dominant and, as a
matter of fact the heavy Higgs H is more important (the Higgs particle
$A$ couples in a pseudoscalar way, which does not lead to coherence).
It is well known that all quark flavors contribute
\cite{Dree}, since the relevant couplings are proportional to
the quark masses.
One encounters in the nucleon not only the usual 
sea quarks ($u {\bar u}, d {\bar d}$ and $s {\bar s}$) but the 
heavier quarks $c,b,t$ which couple to the nucleon via two gluon 
exchange, see e.g. Drees {\it et al}  ~\cite{Dree00} and references
therein.
 
As a result  one obtains an effective scalar Higgs-nucleon
coupling by using  effective quark masses as follows
\begin{center}
$m_u \ra f_u m_N, \ \ m_d \ra f_d m_N. \ \ \  m_s \ra f_s m_N$   
\end{center}
\begin{center}
$m_Q \ra f_Q m_N, \ \ (heavy\ \  quarks \ \ c,b,t)$   
\end{center}
where $m_N$ is the nucleon mass. The isovector contribution is now
negligible. The parameters $f_q,~q=u,d,s$ can be obtained by chiral
symmetry breaking 
terms in relation to phase shift and dispersion analysis.
Following Cheng and Cheng ~\cite{Chen} we obtain:
\begin{center}
$ f_u = 0.021, \quad f_d = 0.037, \quad  f_s = 0.140$ 
\quad  \quad  (model B)   
\end{center}
\begin{center}
$ f_u = 0.023, \quad f_d = 0.034, \quad  f_s = 0.400$ 
\quad  \quad  (model C)   
\end{center}
 We see that in both models the s-quark is dominant.
Then to leading order via quark loops and gluon exchange with the
nucleon one finds:
\begin{center}
\quad $f_Q= 2/27(1-\quad \sum_q f_q)$ , \quad i.e.  
\quad $ f_Q = 0.060$    (model B),   
\quad $ f_Q = 0.040$    (model C)   
\end{center}
 There is a correction to the above parameters coming from loops
involving s-quarks \cite {Dree00} and due to QCD effects. 
 Thus for large $tan \beta$ we find \cite {ref1}:
\begin{center}
\quad $f_{c}=0.060 \times 1.068=0.064,
       f_{t}=0.060 \times 2.048=0.123,
       f_{b}=0.060 \times 1.174=0.070$  \quad (model B)
\end{center}
\begin{center}
\quad $f_{c}=0.040 \times 1.068=0.043,
       f_{t}=0.040 \times 2.048=0.082,
       f_{b}=0.040 \times 1.174=0.047$  \quad (model B)
\end{center}
For a more detailed discussion we refer the reader to 
Refs~\cite{Dree,Dree00}.

\section{Results and Discussion}
\begin{figure}
\includegraphics[height=.3\textheight]{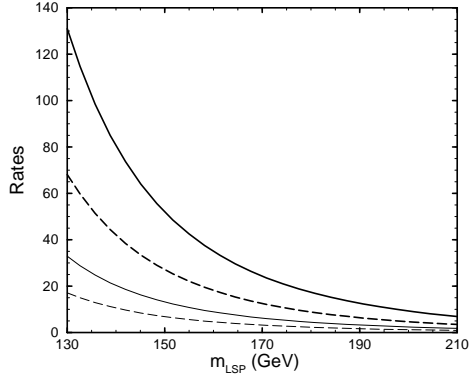}
\caption{The Total detection rate per $(kg-target)yr$ vs the LSP mass
in GeV for a typical solution in our parameter space in the case of 
$^{127}I$ corresponding to  model B (thick line) and Model C (fine 
line). For the definitions see text.
\label{rate}
}
\end{figure}

\begin{figure}
\hspace*{-0.0 cm}
\includegraphics[height=.2\textheight]{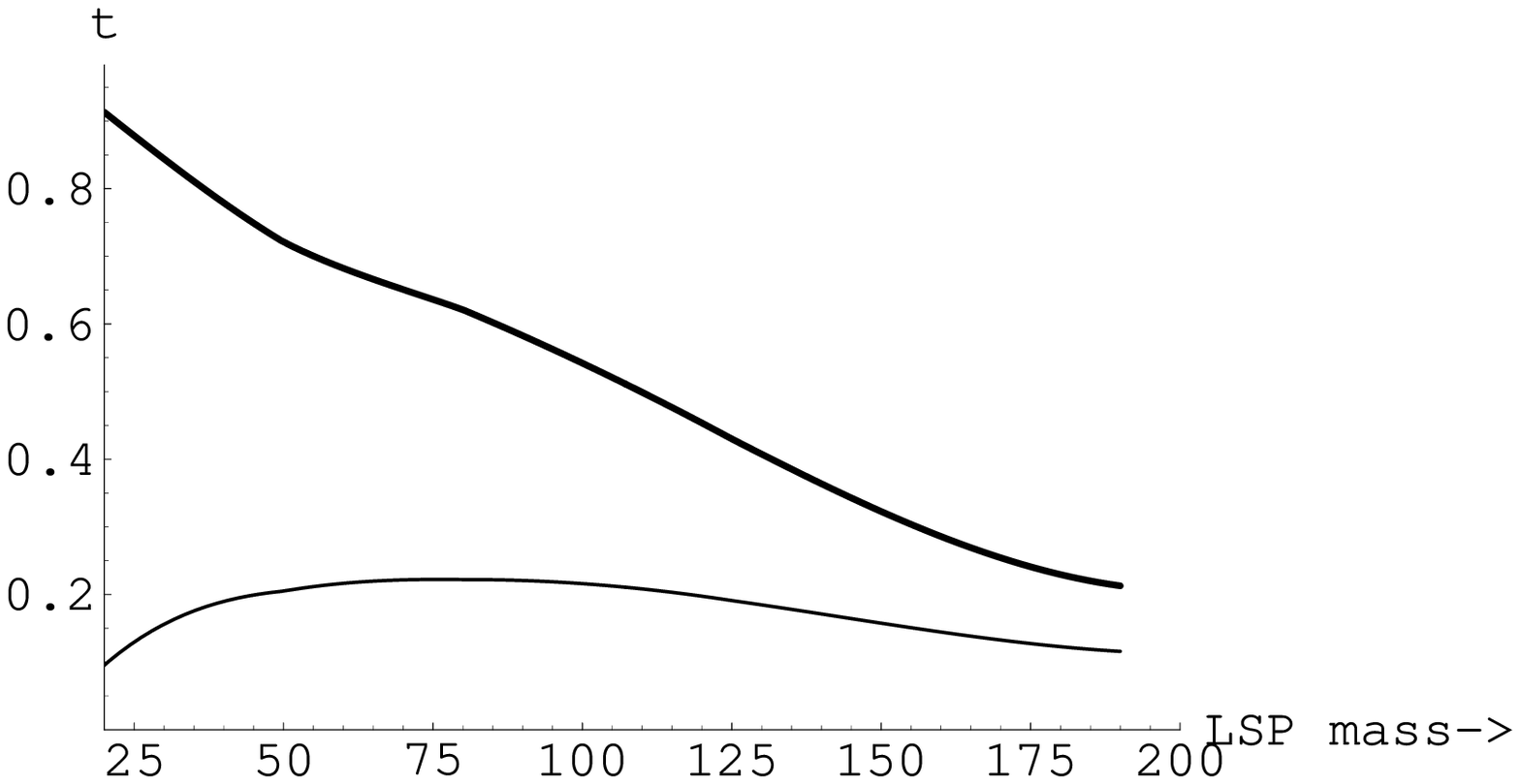}
\includegraphics[height=.2\textheight]{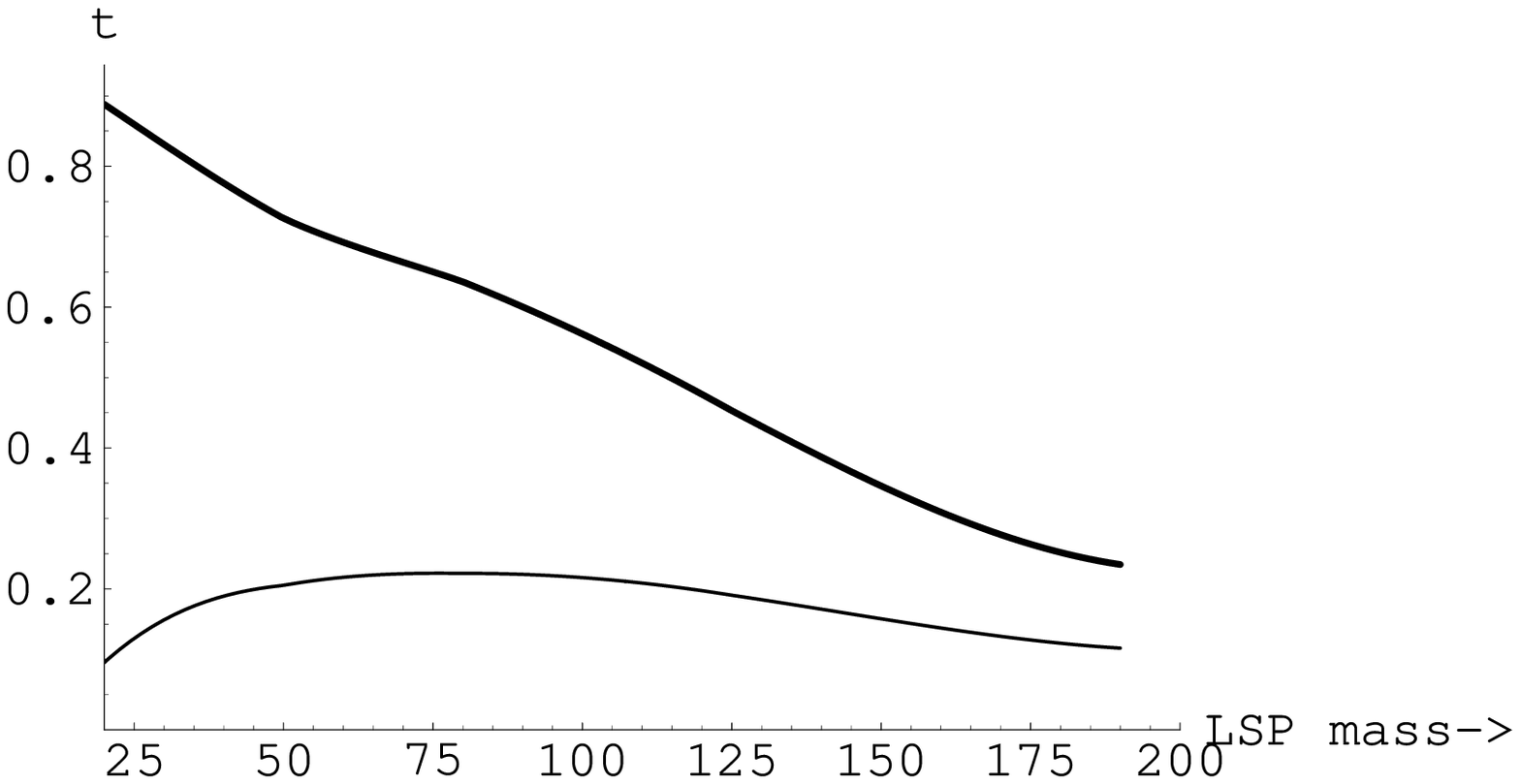}
\caption{
The dependence of the quantity $t$ on the LSP mass
 for the symmetric case ($\lambda=0$) on the left as well as for the
 maximum axial asymmetry ($\lambda=1$) on the right in the case of the
 target $^{127}I$.
For orientation purposes  two  detection cutoff energies are exhibited,
$Q_{min}=0$ (thick solid line) and $Q_{min}=10~keV$ (thin solid line). 
 As expected $t$ decreases as the cutoff energy and/or the LSP 
mass increase. We see that the asymmetry parameter $\lambda$ has little
effect on the un modulated rate. 
\label{fig.t}
}
\end{figure}

\begin{figure}
\includegraphics[height=.3\textheight]{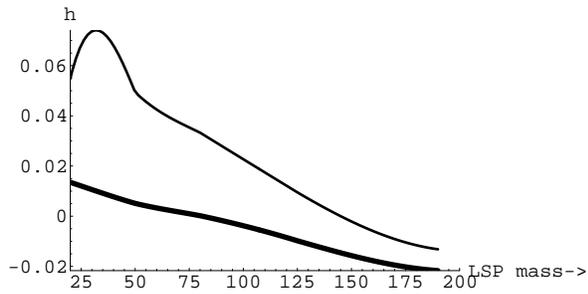}
\caption{ The same as in Fig. \ref{fig.t}
for the modulation with $\lambda=0$. We see that the
modulation is small and decreases with the LSP mass. It even changes sign
for large LSP mass. The
 introduction of a cutoff $Q_{min}$ increases the modulation (at the expense
of the total number of counts).
\label{fig.h0}
}
\end{figure}

\begin{figure}
\includegraphics[height=.3\textheight]{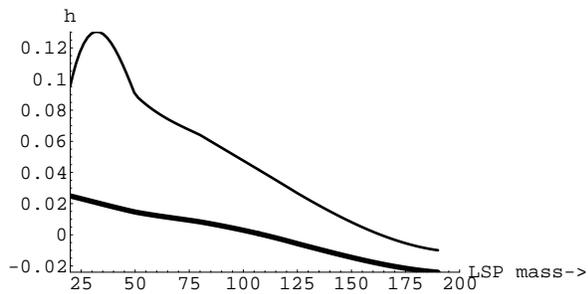}
\caption{ The same as in Fig. \ref{fig.h0} for $\lambda=1$.
We see that the modulation increases
 with the asymmetry parameter $\lambda$.
\label{fig.h1}
}
\end{figure}

The three basic ingredients of our calculation were 
the input SUSY parameters (see sect. 1), a quark model for the nucleon
(see sect. 3) and the velocity distribution combined with the structure of
 the nuclei involved (see sect. 2). we will focus our attention on the
coherent scattering and present results for the popular target $^{127}I$.
We have utilized two nucleon models indicated by B and C which take into
account the presence of heavy quarks in the nucleon. We also considered
energy cut offs imposed by the detector,  
 by considering two  typical cases $Q_{min}=0,~10$ KeV. The thus obtained
results for the un modulated non directional event rates $\bar{R}t$
 in the case
 of the symmetric isothermal model for a typical SUSY parameter choice
\cite{Gomez} are shown in Fig. \ref{rate}. 
  The two relative 
parameters, i.e. the quantities $t$ and $h$, are shown in Fig. \ref{fig.t}
and Figs \ref{fig.h0},\ref{fig.h1} respectively in the case of isothermal
 models.
 The case of non isothermal models, e.g. caustic rings, is more
complicated \cite{Verg01} and it will not be further discussed here.
 
It is instructive to examine the reduction factors along the three axes, i.e 
along $+z,-z,+y,-y,+x$ and $-x$
\cite{Verg00}. Since $f_{red}$ is the ratio of two parameters, its dependence
on $Q_{min}$ and the LSP mass is mild. So we present results for
 $Q_{min}=0$
and give an  average as a function of the LSP mass (see Table 
\ref{table.dir}). As expected the maximum rate is along the sun's direction of
 motion, i.e opposite to its velocity $(-z)$ in the Gaussian distribution
 and $+z$ in the case of caustic rings. In fact we find
that $\kappa(-z)$ is around 0.5 (no asymmetry) and around 0.6 (maximum
asymmetry, $\lambda=1.0$), It is not very different from the 
naively expected $f_{red}=1/( 2 \pi)=\kappa=1$.
The asymmetry
$|R_{dir}(-)-R_{dir}(+)|/(R_{dir}(-)+R_{dir}(+))$ is quite large in the
 isothermal
model and smaller in caustic rings. The rate in the other directions is quite
a bit smaller (see Table \ref{table.dir}).

\begin{figure}
\includegraphics[height=.3\textheight]{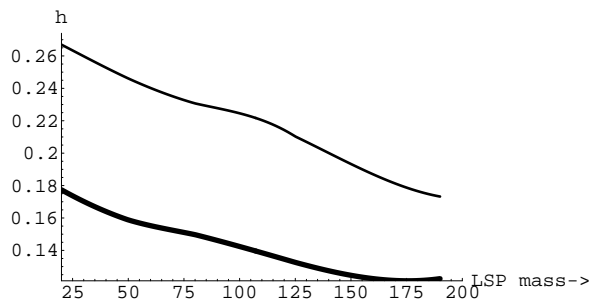}
\caption{ The quantity $h_1-h_2$ in the direction $+y$ for
 $\lambda=0$ (thick line) and $\lambda=1$ thin line. In the $-y$ direction
this quantity is constant and negative, $-0.20$ and $-0.30$ for $\lambda=0$
 and $1$ respectively. As a result the modulation effect is opposite 
(minimum in June the 3nd).
\label{fig.h12}
}
\end{figure}

\begin{figure}
\includegraphics[height=.3\textheight]{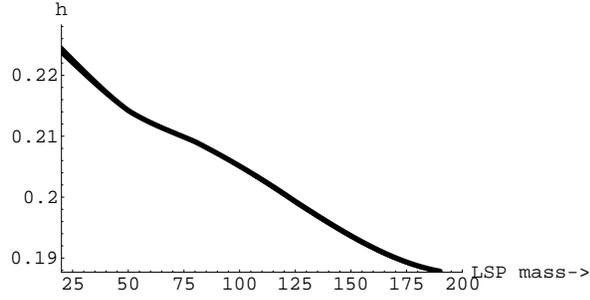}
\caption{ The same as in Fig. \ref{fig.h12} for
the modulation amplitude in the direction $+x$, which is essentially
$h_3$, since$|h_1|<<|h_3|$.  Thus in this case
the maximum occurs around September the 3nd and the minimum 6 months later.
The opposite is true in the $-x$ direction.
\label{fig.h13}
}
\end{figure}

 As we have seen the modulation can be described in terms of the parameters
$h_i,~i=1,2,3$ (see Eq.  (\ref{4.56})). If the observation is done in the
direction opposite to the sun's direction of motion the modulation amplitude
$h_1$ behaves in the same way as the non directional one, namely $h$. It is
instructive to consider directions of observation in the plane perpendicular
 to the sun's
direction of motion ($\Theta=\pi/2$)even though the un modulated rate is reduced
in this direction. Along the $-y$ direction ($\Phi=(3/2)\pi$)
the modulation amplitude $h_1-h_2$ is constant, $-0.20$ and $-0.30$ for
 $\lambda=0,1$ respectively. In other words it large and leads to a maximum
in December. Along the $+y$ direction the modulation is exhibited in Figs
\ref{fig.h12} and \ref{fig.h13}.
\begin{table}[t]  
\caption{
The ratio $\kappa$ of the un modulated directional rate along the three
 directions to the non-directional one: z is in the direction of the sun's
 motion, x is in the radial direction  and x is perpendicular to the axis 
of the galaxy. The asymmetry is also given. $Q_{min}=0$ was assumed.
}
\label{table.dir}
\begin{tabular}{lrrrrrrr}
& & & & & & &    \\
&  & \multicolumn{3}{c|}{isothermal} & \multicolumn{3}{c|}{caustic rings}  \\ 
 &   &  &  & & & &     \\
  $\lambda$ & dir. & +  & - &asym&+&-&asym\\
\hline
    &   &  &  &      \\
0    & z   & 0.02  & 0.50  &  0.92&0.75&0.25&0.50\\
0    & y   & 0.16  & 0.16  &  0&0.22&0&1.00\\
0    & x  &  0.16  & 0.16  &  0&0.37&0.24&0.21\\
1    & z   & 0.04  & 0.58  &  0.90&-&-&- \\
1    & y   & 0.12  & 0.12  &  0&-&-&-\\
1    & x  &  0.17  & 0.17  &  0&-&-&-\\
\end{tabular}
\end{table}
\section{Conclusions}
In the present paper we have discussed the parameters, which describe
the event rates for direct detection of SUSY dark matter.
Only in a small segment of the allowed parameter space the rates are above
 the present experimental goals.
 We thus looked for
characteristic experimental signatures for background reduction, i.e.
a) Correlation of the event rates
with the motion of the Earth (modulation effect) and b)
 the directional rates (their correlation  both with the velocity of the sun
 and that of the Earth.)

A typical graph for the total un modulated rate is shown Fig. \ref{rate}. 
The relative parameters $t$ and $h$ in the case of non directional experiments
are exhibited in Fig. \ref{fig.t} and  Figs \ref{fig.h0} and \ref{fig.h1}.
We must emphasize that these two graphs do not contain the entire dependence on
the LSP mass. This is due to the fact that there is the extra factor $\mu_r^2$
in Eq. ({2.10}) but mainly due to the fact that the nucleon cross section
 depends on the LSP mass. Fig
 \ref{fig.t} and  Figs \ref{fig.h0} and \ref{fig.h1}.
 were obtained for
 the scalar interaction, but do not expect a very different behavior in the
case of the spin contribution. The overall spin contribution,
however, is going to be very different, but such considerations are beyond
the goals of the present paper.
 We should also mention that in the non directional
experiments the modulation $2h_1$ is small, .i.e. for $\lambda=0$ less than
 $4\%$ for
$Q{min}=0$ and $12\%$ for $Q{min}=10~KeV$ (at the expense of the total
number of counts). For $\lambda=1$ there in no change for
$Q_{min}=0$, but it can go as high as $24\%$ for $Q_{min}=10~KeV$. 

For the directional rates It is instructive to examine the reduction factors
 along the three axes, i.e 
along $+z,-z,+y,-y,+x$ and $-x$. These depend on the 
nuclear parameters, the reduced mass, the energy cutoff $Q_{min}$ and 
 $\lambda$
\cite{Verg00}. Since $f_{red}$ is the ratio of two parameters, its dependence
on $Q_{min}$ and the LSP mass is mild. So we present results for $Q_{min}=0$
and give their average as a function of the LSP mass (see Table 
\ref{table.dir}). As expected the maximum rate is  along the sun's direction of
 motion, i.e opposite to its velocity $(-z)$ in the Gaussian distribution
 and $+z$ in the case of caustic rings. In fact we find
that $\kappa(-z)$ is around 0.5 (no asymmetry) and around 0.6 (maximum
asymmetry, $\lambda=1.0$). It is not very different from the 
naively expected $f_{red}=1/( 2 \pi)=\kappa=1$.
The asymmetry along the sun's direction of motion,
$asym=|R_{dir}(-)-R_{dir}(+)|/(R_{dir}(-)+R_{dir}(+))$ is quite characteristic,
 i.e.  quite large in the isothermal
models and smaller in caustic rings. The rate in the other directions is quite
a bit smaller (see Table \ref{table.dir}). But, as we have seen, in the plane
perpendicular in the sun's velocity we have very large modulation, which may
 overcompensate for the large reduction factor. It is interesting to note that
 the dependence on the phase of the Earth $\alpha$ changes substantially with
the direction of observation.

 In conclusion in the case of directional un modulated rates we expect
 characteristic and pretty much
 unambiguous correlation with the motion on the sun, which can be explored by
 the experimentalists. The reduction factor in the direction of the motion
of the sun is approximately only $1/(4\pi)$ relative to the non directional
experiments.  In the plane perpendicular to the motion of the sun we expect
interesting modulation signals, but the reduction factor becomes worse.
A more complete discussion will be given elsewhere \cite{JDVb02}.

\par
This work was supported by the European Union under the contracts 
RTN No HPRN-CT-2000-00148 and TMR 
No. ERBFMRX--CT96--0090.

\def\ijmp#1#2#3{{ Int. Jour. Mod. Phys. }{\bf #1~}(#2)~#3}
\def\pl#1#2#3{{ Phys. Lett. }{\bf B#1~}(#2)~#3}
\def\zp#1#2#3{{ Z. Phys. }{\bf C#1~}(#2)~#3}
\def\prl#1#2#3{{ Phys. Rev. Lett. }{\bf #1~}(#2)~#3}
\def\rmp#1#2#3{{ Rev. Mod. Phys. }{\bf #1~}(#2)~#3}
\def\prep#1#2#3{{ Phys. Rep. }{\bf #1~}(#2)~#3}
\def\pr#1#2#3{{ Phys. Rev. }{\bf D#1~}(#2)~#3}
\def\np#1#2#3{{ Nucl. Phys. }{\bf B#1~}(#2)~#3}
\def\npps#1#2#3{{ Nucl. Phys. (Proc. Sup.) }{\bf B#1~}(#2)~#3}
\def\mpl#1#2#3{{ Mod. Phys. Lett. }{\bf #1~}(#2)~#3}
\def\arnps#1#2#3{{ Annu. Rev. Nucl. Part. Sci. }{\bf
#1~}(#2)~#3}
\def\sjnp#1#2#3{{ Sov. J. Nucl. Phys. }{\bf #1~}(#2)~#3}
\def\jetp#1#2#3{{ JETP Lett. }{\bf #1~}(#2)~#3}
\def\app#1#2#3{{ Acta Phys. Polon. }{\bf #1~}(#2)~#3}
\def\rnc#1#2#3{{ Riv. Nuovo Cim. }{\bf #1~}(#2)~#3}
\def\ap#1#2#3{{ Ann. Phys. }{\bf #1~}(#2)~#3}
\def\ptp#1#2#3{{ Prog. Theor. Phys. }{\bf #1~}(#2)~#3}
\def\plb#1#2#3{{ Phys. Lett. }{\bf#1B~}(#2)~#3}
\def\apjl#1#2#3{{ Astrophys. J. Lett. }{\bf #1~}(#2)~#3}
\def\n#1#2#3{{ Nature }{\bf #1~}(#2)~#3}
\def\apj#1#2#3{{ Astrophys. Journal }{\bf #1~}(#2)~#3}
\def\anj#1#2#3{{ Astron. J. }{\bf #1~}(#2)~#3}
\def\mnras#1#2#3{{ MNRAS }{\bf #1~}(#2)~#3}
\def\grg#1#2#3{{ Gen. Rel. Grav. }{\bf #1~}(#2)~#3}
\def\s#1#2#3{{ Science }{\bf #1~}(19#2)~#3}
\def\baas#1#2#3{{ Bull. Am. Astron. Soc. }{\bf #1~}(#2)~#3}
\def\ibid#1#2#3{{ ibid. }{\bf #1~}(19#2)~#3}
\def\cpc#1#2#3{{ Comput. Phys. Commun. }{\bf #1~}(#2)~#3}
\def\astp#1#2#3{{ Astropart. Phys. }{\bf #1~}(#2)~#3}
\def\epj#1#2#3{{ Eur. Phys. J. }{\bf C#1~}(#2)~#3}

\end{document}